\documentclass[prd,twocolumn,superscriptaddress,preprintnumbers,nofootinbib,floatfix]{revtex4-2}
\usepackage{epsfig}
\usepackage{graphicx}
\usepackage[english]{babel}
\usepackage{hyphenat}
\usepackage{amsmath}
\usepackage{amssymb}
\usepackage{mathtools}
\usepackage{mathrsfs}
\usepackage{slashed}
\usepackage{epstopdf}
\usepackage{xcolor}
\usepackage{braket}
\usepackage{booktabs}
\definecolor{lcolor}{rgb}{0.,0.0,0.}
\definecolor{citcolor}{rgb}{0,0.,0.5}
\usepackage[breaklinks,colorlinks,urlcolor=blue,citecolor=blue,linkcolor=blue]{hyperref}
\usepackage{multirow}
\usepackage{ltablex}
\usepackage{soul}
\usepackage{wrapfig}

\newcommand{\beq}{\begin{eqnarray}}
\newcommand{\eeq}{\end{eqnarray}}

\newcommand{\bem}{\begin{multline}}
\newcommand{\eem}{\end{multline}}
\newcommand{\beg}{\begin{gather}}
\newcommand{\eeg}{\end{gather}}

\newcommand{\nn}{\nonumber\\}

\newcommand{\ben}{\begin{eqnarray*}}
\newcommand{\een}{\end{eqnarray*}}

\newcommand{\secn}[1]{Section~1}
\newcommand{\appn}[1]{Appendix~1}

\long\def\comment#1{ }

\def\Tr{\text{Tr}}

\def\and{\quad\text{and}\quad}

\def\0{{\boldsymbol 0}}

\def\0{{\boldsymbol 0}}

\begin{document}

\title{Real-time simulation of jet energy loss and entropy production \\ in high-energy scattering with matter}

\author{Jo\~{a}o Barata}
\email[Email: ]{joao.lourenco.henriques.barata@cern.ch}
\affiliation{CERN, Theoretical Physics Department, CH-1211 Geneva 23, Switzerland}

\author{Enrique Rico}
\email[Email: ]{enrique.rico.ortega@cern.ch}
\affiliation{CERN, Theoretical Physics Department, CH-1211 Geneva 23, Switzerland}

\preprint{CERN-TH-2025-019}

\begin{abstract}
In analogy to high-energy nuclear scattering experiments, we study a real-time scattering process between a propagating state and a dense target in $1+1$-d massive QED. In our setup, we identify three distinct regimes that qualitatively characterize the evolution: for a dilute medium, the incoming probe state evolves nearly ballistically; in an intermediate setting, it traverses the matter, locally exciting it; and for dense targets, one approaches a black-disk limit, where the matter acts as a strong wall potential. We find evidence that the probe's energy loss rate scales linearly with the path length in the medium, and we study how the entanglement entropy reveals the mixing between the probe and medium states. With the goal of one day replicating high-energy nuclear experiments in quantum devices, we briefly discuss how the current tensor network-based simulations can be translated to a quantum simulator.

\end{abstract} 

\maketitle


\section{Introduction}

High-energy nuclear scattering experiments probe small-scale structures inside QCD matter, revealing novel properties of the theory. They have led to the discovery of primordial matter states, such as the quark-gluon plasma (QGP) produced at the LHC and RHIC  -- see~\cite{Busza:2018rrf} for an overview -- and to the construction of a tomographic picture of hadrons' inner structure, see e.g.~\cite{Gao:2017yyd,Boussarie:2023izj}, which will continue to be imaged in the future EIC accelerator at BNL~\cite{AbdulKhalek:2022hcn}. On the theoretical side, there is a vast dedicated literature explaining the phenomena observed in these experiments, mostly starting from a perturbative QCD picture, see e.g.~\cite{Collins:2011zzd,Kovchegov:2012mbw,Ellis:1996mzs}, which has yielded an impressive agreement with the experimental data for systems such as electron-positron, electron-proton, and proton-proton collisions, where the final state particle multiplicities are not too high. Nonetheless, our first-principle understanding of the underlying dynamics is considerably less developed for bigger collisional systems, such as proton-nucleus or nucleus-nucleus collisions, or when focusing on low energy properties of smaller systems. This results from the high degree of non-linearity and the relevance of non-perturbative effects in those scenarios. Even though effective descriptions of the dynamical properties of larger collisional systems ~\cite{Gale:2013da,Gelis:2010nm} and of the low energy features~\cite{Andersson:1997xwk,Winter:2003tt} can be constructed, an overarching framework, with a clear connection to first-principle quantum field theory (QFT), is still lacking.

\begin{figure}[h]
    \centering
    \includegraphics[width=1\columnwidth]{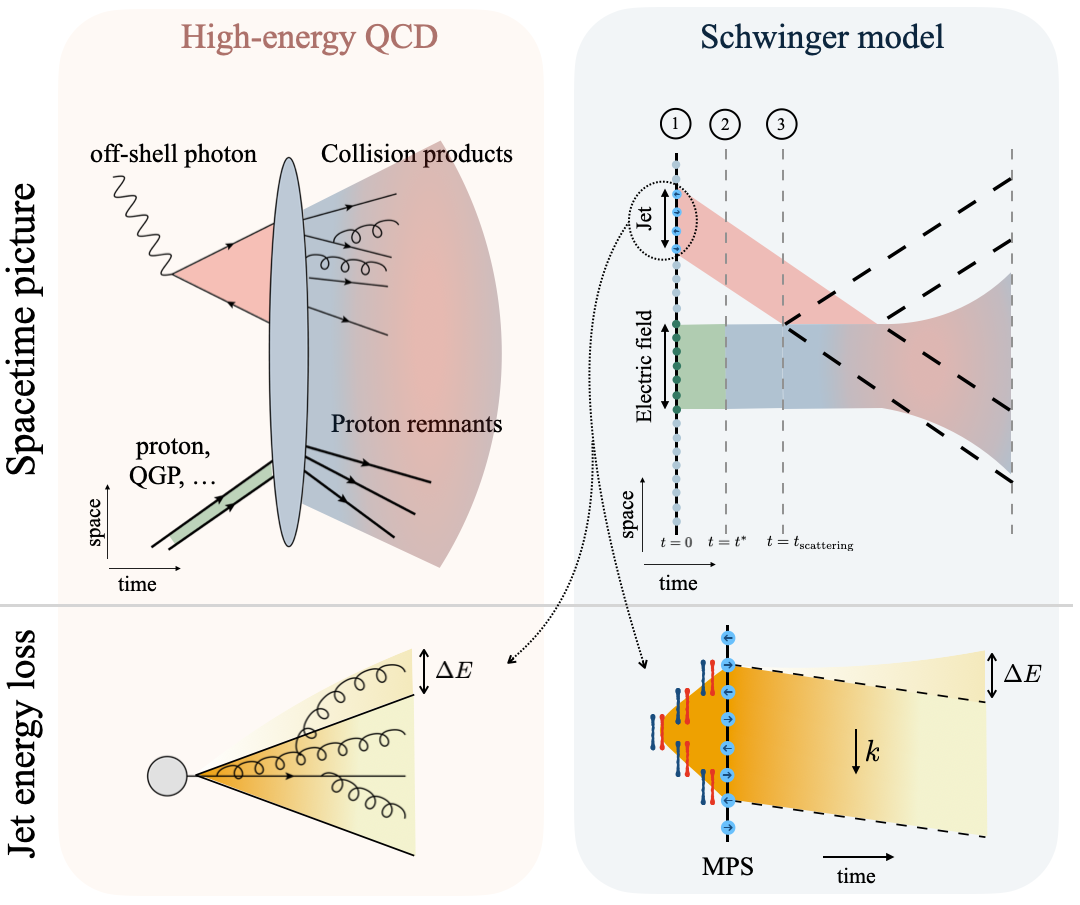}
    \caption{\textbf{Top Left:} Schematic representation of a collision between an off-shell photon state and target proton/QGP state in the high-energy QCD context. Here, the initial photon state can fluctuate to a long-lived zero charge multi-particle state, the leading order one being a quark anti-quark pair, which interacts with the target at rest. The vertical blue oval denotes the interaction. After the interaction, the state can fluctuate to various multi-particle final states. \textbf{Top Right:} An equivalent diagram in the $1+1$-d Schwinger model, representing the main steps followed in the simulation protocol. In (1) we schematically represent the initial state on the computational lattice, while (2) and (3) denote the initial free propagation of the system and the start of interactions, respectively. \textbf{Bottom Left:} Diagrammatic representation of a contribution to jet energy loss in QCD. Here a jet, i.e. a multi-particle final state reconstructed within some spatial region in the detector, emits a gluon outside of the jet cone (region in between black lines), which leads to the depletion of the total energy of the state by an amount $\Delta E$. \textbf{Bottom Right:} An equivalent interpretation of energy loss in the lattice simulation. We first schematically represent the initial jet state (red and blue lines are the electric fluxes) in the strong coupling regime and its energy distribution in yellow. The natural evolution of the state leads to part of its energy dragging behind, which can be analogously interpreted as a $\Delta E$ energy loss.}
    \label{fig:cartoon}
\end{figure}

One route towards addressing some of these aspects would require replicating these experiments from first-principles (lattice) QFT considerations. However, due to the prevalence of severe noise problems, traditional Euclidean QFT methods can not help build a real-time picture of high-energy nuclear scattering events. In this respect, the development of new quantum technologies, such as quantum computers and simulators, can decisively change this paradigm, since they might in the future allow for long-time and large-scale lattice computations in real-time signature, see~\cite{DiMeglio:2023nsa,Bauer:2022hpo,Halimeh:2023lid} for a summary of recent related efforts. Nevertheless, our ability to efficiently perform real-time simulations of scattering events in gauge theories such as QCD remains limited. Furthermore, preparing the initial state of the collision is expected to be as challenging as solving the real-time problem itself for processes involving nuclear targets.

Despite these limitations, in the last decade there has been a large interest and progress in describing real-time non-perturbative dynamics of simpler lower dimensional (gauges) theories, serving as a benchmark for future calculations, see e.g.~\cite{ Hebenstreit:2014rha,Barata:2023jgd,Banerjee:2012pg,Angelides:2025hjt,Calajo:2024qrc,Magnifico:2019kyj,Ale:2024uxf,Jha:2023ecu,Shaw:2020udc,Charles:2023zbl,Carena:2022kpg,Kurkcuoglu:2024cfv,Farrell:2024fit,Zache:2021ggw,Ott:2020ycj,Spitz:2018eps,Ikeda:2024rzv,Grieninger:2024cdl,ARahman:2022tkr,ARahman:2022tkr,Mueller:2024mmk,Davoudi:2020yln,Janik:2025bbz}. In particular, the real-time dynamics of scattering processes involving simple states, such as \textit{electrons} and \textit{mesons}, in $1+1$-d QFTs has been extensively studied. For example, Ref.\cite{Pichler:2015yqa} provided a first study of scattering between meson wave-packets in $1+1$-d QED, i.e. the Schwinger model, leading to a direct observation of entropy production and its connection to the entanglement between the meson states. More recent studies in the same model have expanded on these findings, see e.g.~\cite{Papaefstathiou:2024zsu,Zemlevskiy:2024vxt,Belyansky:2023rgh}, highlighting the role quantum effects can play in such events.\footnote{See e.g.~\cite{Farrell:2024mgu,Bennewitz:2024ixi,Davoudi:2024wyv,Belyansky:2023rgh,Jordan:2011ci,Jordan:2012xnu,Rigobello:2021fxw,Briceno:2023xcm,Ale:2024uxf, Milsted:2020jmf} for more related works.} Nonetheless, less attention has been given to scattering processes involving a \textit{single-particle} or simple \textit{bound} (probe) state and a more complex \textit{composite} (matter) state. These events allow us to explore new questions, such as how the matter imprints itself onto the probe state, and have direct analogs in high-energy QCD experiments.
 
In this work, we take a first step in this direction by studying a scattering process between a (nearly bound) \textit{meson} state and a compact \textit{matter} state in the Schwinger model. This is illustrated in Fig.~\ref{fig:cartoon} (\textbf{top right}), where the incoming probe scatters off a region initially populated by a strong electric field. For reasons we motivate below, and in analogy to QCD language, we shall refer to the probe as a jet state and the compact state where the electric field is inserted as the matter target. In Fig.~\ref{fig:cartoon} (\textbf{top left}) we show the analogous setup in QCD, where an off-shell chargeless initial state made of two quarks scatters off a resting QCD matter state, as is typically found in deep inelastic scattering and related experiments, see e.g.~\cite{Gribov:1969zz,Bjorken:1973gc,Frankfurt:1988nt,Gelis:2010nm}, or an idealized heavy ion collision experiment, where external probes could be prepared, see e.g.~\cite{Apolinario:2022vzg,Cao:2020wlm}. After the probe-matter interaction, a complex multi-particle final state can be produced, as indicated in the diagram. As we discuss below, the detailed dynamics of this experiment are rather rich, encompassing both a regime with ballistic propagation of the initial jet state, an intermediate scenario where the matter modifies the jet, and an extreme regime where the jet scatters off an opaque target, characterized by an initial strong electric field. Finally, one can access more information in these simulations than equivalent QCD experiments. While in the former, expectation values of operators can be computed between any combination of spacetime points, in the latter one only has access to the particle distributions measured by the detectors.

This work is divided as follows. In section~\ref{section:model} we introduce the Schwinger model, followed by a discussion of the simulation protocol in section~\ref{sec:protocol}. We present the numerical results in section~\ref{sec:results}. We briefly discuss how these studies could be carried out in quantum simulators, going beyond the present tensor network-based study, in section~\ref{sec:synth_platt}, before concluding the paper in section~\ref{sec:conclusion}.

\section{Lattice Schwinger model}\label{section:model}

The theory of quantum electrodynamics in $1+1$-d dimensions was first studied in depth by Schwinger~\cite{Schwinger:1951nm}, pointing out the remarkable phenomena of pair production out of the vacuum applying high-intensity external electric fields. In the coming years, further interesting aspects of the theory were discovered and studied at length, such as e.g. the phenomena of Abelian bosonization and the duality to Sine-Gordon theories~\cite{Coleman:1974bu}, and the low energy properties at strong coupling, leading to important developments in the understanding of other QFTs, see e.g.~\cite{Susskind:1976jm, Kogut:1976rs}. As such, the Schwinger model is the prime testbed to explore complex phenomena relevant to more realistic theories of Nature.

In the Hamiltonian formalism, the theory can be conveniently studied by latticing the spatial dimension. In the $A^0=0$ gauge, the continuum Hamiltonian reads~\cite{Schwinger:1951nm}
\begin{align}\label{eq:H_Schwinger}
	H&=  \int dx \, \frac{1}{2 } E^2(x) \nn 
 &+ \bar  \psi(x) (-i \gamma^1 \partial_1 +g \gamma^1 A_1(x) +m ) \psi(x)\, ,
\end{align}
where $\psi$ denotes the single flavor two-component spinor field with mass $m$, $g$ is the coupling constant, and $A^1$ is the non-vanishing component of the gauge field, which can be traded for the electric field $E=-\partial_t A^1$. In addition, the equations of motion of the gauge degrees of freedom give rise to a constrain equation (Gauss' law), $\partial_x E=g  \psi^\dagger \psi$, which allows the explicit integration of $E$, leading to a theory written solely in terms of fermionic degrees of freedom with a non-local fermionic potential. 

The lattice theory can be directly constructed from Eq.~\eqref{eq:H_Schwinger}, following, for example, the Kogut-Susskind prescription for fermions~\cite{Susskind:1977lf,Kogut:1974ag}. This maps the two-component spinor $\psi$ to single component operators $\chi$:
\begin{align}\label{eq:help_1}
   \psi(\tilde n )  \to \frac{1}{\sqrt{a}}\begin{pmatrix}
		\chi_{2 \tilde n}\\
		\chi_{2 \tilde  n-1}
	\end{pmatrix}\, ,
\end{align}
with $a$ the lattice spacing, the subindex indicates the location on the lattice. The staggered fermion prescription effectively extends the physical lattice naively obtained from the continuum by a factor of two. In what follows we shall use un-tilde symbols to refer to the staggered lattice, which we assume has $N$ sites. The reduced electric field $ L\equiv  E/g$ can be integrated out using the local Gauss' law generator
 \begin{align}
G_n&=\delta L_n-  \Bigg( \chi_n^\dagger \chi_n -\frac{1}{2}(1-(-1)^n)\Bigg)\, ,
\end{align}
where $\delta L_n\equiv  L_n-L_{n-1}$. In the charge zero sector, one has that any physical state satisfies $G|\psi\rangle =0$. 

The model can be further mapped to an equivalent spin-chain. To that end, one performs a Jordan-Wigner transform (JWt)~\cite{Jordan:1928wi} to the single component spinors
\begin{align}
	&\chi_n =      \prod_{ k< n}[(-i)\sigma^z_k ] \sigma^-_n, \quad \chi^\dagger_n =     \prod_{ k< n}[i\sigma^z_k ] \sigma^+_n\, ,
\end{align}
where $\sigma^{x,y,z,+,-}_n$ denote the standard Pauli matrices acting on the site $n$. We define $\sigma^{\pm} = (\sigma^x\pm i \sigma^y)/2$. Solving for Gauss' law with open boundary conditions (with vanishing electric and fermion fields at the edges) one has~\cite{Hamer:1997dx}:
\begin{align}\label{eq:Spin_Hamiltonian}
	H(t) &= 	\frac{g^2a}{2} \sum_{n=1}^{N-1} \left[\frac{1}{2}\sum_{k=1}^n (\sigma^z_k+(-1)^k)  + \alpha_{\rm ext}(t,n)\right]^2  \nn 
    &+\sum_{n=1}^{N} m  (-1)^n \frac{\sigma^z_n}{2} \nn
	&+\frac{1}{2a} \sum_{n=1}^{N-1}  \sigma^+_n \sigma^-_{n+1} + \sigma^+_{n+1}  \sigma^-_n  \, ,
\end{align}
where the $\gamma$ matrices are represented in the basis $\gamma^0=\sigma^z$, $\gamma^1 = -i \sigma^y$, and $\gamma^5=-\sigma^x$. In Eq.~\eqref{eq:Spin_Hamiltonian} we have added the term $\alpha_{\rm ext}$ in the gauge sector to account for the possibility of having an external non-dynamical electric field, as it will be relevant below. In particular, in this work, we shall assume this takes the continuum form
\begin{align}
   \alpha_{\rm ext}(t,x) = -|Q| \Theta(t^*-t) \Theta(x^-<x<x^+) \, ,
\end{align}
where $|Q|$ can be understood as the charge of the non-dynamical external fermions generating the field, $t^*$ sets the sudden turn-off time of the field, and $x^\pm$ are the spatial edges of the field's domain, see Fig.~\ref{fig:cartoon}.

Finally, in what follows we track the evolution of local observables over the entire simulation time. We consider the following set of observables:
\begin{itemize}
    \item Local Condensate: $a \langle \bar \psi \psi\rangle_n = (-1)^n \langle \sigma^z_n\rangle$, where we dropped overall constant factors and terms that vanish under the sum in $n$.
    \item Local Electric field: $L(n) = \frac{1}{2} \sum_{k=1}^n \left((-1)^k + \sigma_k^z\right)$.
    \item Local Energy and jet energy loss: We define the local energy at a site $n$ in terms of the local Hamiltonian $H(n)$, such that 
    \begin{align}
        \mathcal{E}(n) = \langle\psi(t)| H(n) |\psi(t)\rangle -\langle \Omega | H(n) |\Omega \rangle \, ,
    \end{align}
    where $|\Omega \rangle$ denotes the vacuum state and $|\psi(t)\rangle$ is full system's state at time $t$.  This allows us to define the local energy loss rate in the vacuum
    \begin{align}
       \frac{d\mathcal{E}^{\rm jet, \, vac}}{ds} &=  \lim_{\Delta t \to 0 }  \frac{ \mathcal{E}^{\rm in}(t+\Delta t)-\mathcal{E}^{\rm in}(t)  }{\Delta t}\, , \nn 
    \mathcal{E}^{\rm in}(t)&=\sum\limits_{n^-<n<n^+} \mathcal{E}(n,t)\, ,
    \end{align}
     we note that the infinitesimals $ds$, and $dt$ are directly proportional for linear trajectories and we shall drop the overall constant. The sum over energies is time-dependent, i.e. $n^\pm = n^\pm(t)$, and $n^\pm$ define the boundaries of the jet state as it evolves. In the case where we have a background medium, we further define the purely medium-induced jet energy loss as 
    \begin{align}
        \left|\frac{d\mathcal{E}^{\rm jet, \, med}}{ds}\right| = \left|\frac{d\mathcal{E}^{\rm jet, \, full}}{ds}- \frac{d\mathcal{E}^{\rm jet , \, vac}}{ds} \right|\, ,
    \end{align}
     the vacuum term is defined above and the full term is the quantity computed in the full scattering simulation after removing the energy stored in the injected matter state. This can be computed by performing the simulation without the jet.
    \item Local entanglement entropy: We compute the local entropy of entanglement, bi-partitioning the system into a subsystem made of the qubits up to site $n$, and another subsystem with the qubits to the right. Defining the reduced density matrix of the system up to site $n$ as $\rho^l(n)$, one has
    \begin{align}
        S(n) = - \Tr \, \rho^l(n) \log \rho^l(n) \, .
    \end{align}
     In what follows, we always compute the entropy variation subtracting the entropy of the vacuum.
\end{itemize}

\section{Scattering protocol}\label{sec:protocol}

 On the top right-hand side of Fig.~\ref{fig:cartoon} we present a diagrammatic depiction of the simulation protocol followed in this work. This is split into the following stages: 
 
\begin{enumerate}

\item \textbf{Initial state preparation (1):} We start by preparing the ground state of the Schwinger model at finite $m/g\ll 1$ while setting $\alpha_{\rm ext.}=0$. We work close to the strong coupling limit, i.e. $(a g)^{-1}\to 0 $, where in the asymptotic limit the vacuum state is that of the Ising model~\cite{Banks:1975gq}, i.e. $|\Omega\rangle = | \uparrow \downarrow \uparrow \downarrow\cdots \rangle$.
    
\item \textbf{Initial quench, (1) to (2):} We then quench the ground state by inserting on the top half of the lattice the jet state, and below we inject a non-vanishing $\alpha_{\rm ext.}$ at $t>0^+$. 
    
The jet state can be (approximately) prepared close to the strong coupling limit. In that case, one can show that the lowest lying state above the vacuum is a vector meson~\cite{Banks:1975gq}: 
\begin{align}\label{eq:1V}
    | 1_V\rangle = \frac{1}{\sqrt{N }}\sum_n \left( \sigma^+_{n}\sigma^-_{n+1}- \sigma^+_{n+1}\sigma^-_{n}\right) |\Omega\rangle \, .
\end{align}
When close, but not exactly at infinite coupling, one expects that localized excitations with a nearly definite momentum quantum number can be constructed. This observation has been previously successfully applied in~\cite{Papaefstathiou:2024zsu,Pichler:2015yqa}. In particular, here we use the strategy utilized in~\cite{Papaefstathiou:2024zsu} to create a meson wave-packet as the initial condition for the jet:
\begin{align}\label{eq:jet_init}
|\phi_j\rangle &\propto \sum_{n= n ^-}^{n^+} \phi_n(j) e^{-i n k} \nn 
&\times \left( \sigma^+_{n}\sigma^-_{n+1}- \sigma^+_{n+1}\sigma^-_{n}\right) |\Omega'\rangle \, ,
\end{align}
where $|\Omega' \rangle$ should be understood as the finite $m/g$ vacuum, and $|\phi_j\rangle$ is a Gaussian wave-packet centered around $j$th site, with dispersion $\sigma$ and momentum $|k|=a^{-1}$. The energy profile of this state is denoted in Fig.~\ref{fig:cartoon} (\textbf{bottom right}), where the red and green lines represent the respective electric fluxes.

We note that $|\phi_j\rangle$ is not an eigenstate of $H$. As a result, the natural evolution of this state leads to the spreading of the wave packet. Below, we show that, for the simulation times considered, this effect is not dominating. Nonetheless, we note that the spreading of the state leads to outward energy flow, matching the QCD picture for the highly virtual initial states of jets. For this reason, it is reasonable to interpret the prepared state as analogous to a QCD jet, rather than a meson state. Of course, such a statement should be taken at a qualitative level, since the Schwinger model and QCD have drastically different features.

After preparing this initial state, we evolve the full system with the electric field turned on, till a time $t^*$. At the same time, the jet state moves towards the region of the electric field's support before scattering off the target. The evolution in the electric field region is performed using a smaller fermion mass, thus allowing for a larger particle production rate; this initial quench plays no critical role and represents a particular choice to prepare the target.

\item \textbf{Removing the external charges, (2) to (3):} At the time $t^*$, we turn off the external field, and evolve the system with the free Hamiltonian. We note that the state at $t=t^*$ can be understood as the true initial state of the collision, while the previous steps are just introduced to excite the region populated by the medium.

\item \textbf{Scattering and measurement:} After the initial state is prepared we evolve the full system for a sufficiently long time to allow the jet to scatter on the matter state, such that the collision products have enough time to emerge from the interaction. We monitor the dynamics of the collision by performing local measurements throughout the system's evolution. Of course, in a real high-energy nuclear experiment, such a procedure is impossible, and one is constrained to making measurements at some late time $t_{\rm measurement} \gg t_{\rm scattering}\gg t^*$. 

\end{enumerate}

The numerical results of this simulation protocol are shown in the following section. All simulations are performed using the tensor network software package {\tt{iTensor}}~\cite{itensor}, with the ground state of the theory being prepared using the built-in density matrix renormalization group algorithm (DMRG)~\cite{PhysRevB.48.10345,PhysRevLett.69.2863}. The time evolution is performed with the time-dependent variational principle algorithm (TDVP)~\cite{Haegeman:2011zz,PhysRevB.94.165116}. The states are mapped to a matrix product state (MPS) topology. 

We work in the lattice model with $a=1$, and we do not attempt to extrapolate to the continuum. Working close to the strong coupling limit, we take $g a=2$, and $m a=0.1$. To excite the electric field, we locally take the mass in the region before $t^*$ to be $m a =0.0001$, and we take $0\leq |Q|\leq1$; while we have explored higher values of the charge, it does not lead to any new qualitative behavior. We work on a lattice with $N=60$, while the initial injected electric field is located between sites 27 and 33. Finally, we select $t^* a^{-1}=6$, while the total simulation time is $t_{\rm measurement} a^{-1}=160$. The scattering time is of $\mathcal{O}(60 \, a)$, and thus we satisfy the hierarchy $t_{\rm measurement} \gg t_{\rm scattering}\gg t^*$. We also note that the initial collisional states are prepared sufficiently far apart, so they do not interfere. 

Throughout the time evolution in the simulation, and for the most complex case, the maximal bond dimension $D$ was $D\sim\mathcal{O}(460)$, below the maximally allowed bond dimension of $D=800$. We also checked that changing the convergence criteria/tolerance for the TDVP algorithm did not qualitatively affect the results. 

\begin{figure*}[t!]
    \centering
    \includegraphics[width=.3\textwidth]{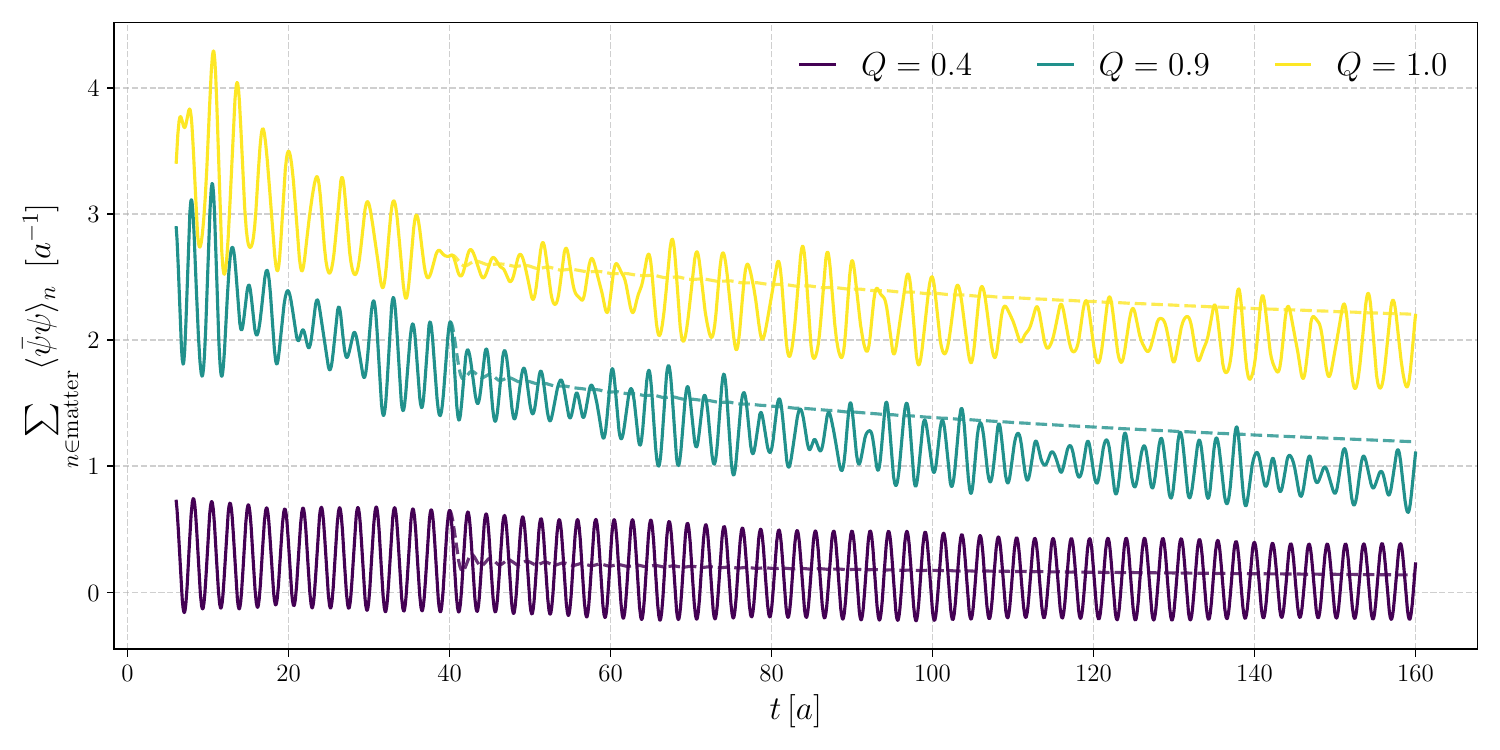}
    \includegraphics[width=.3\textwidth]{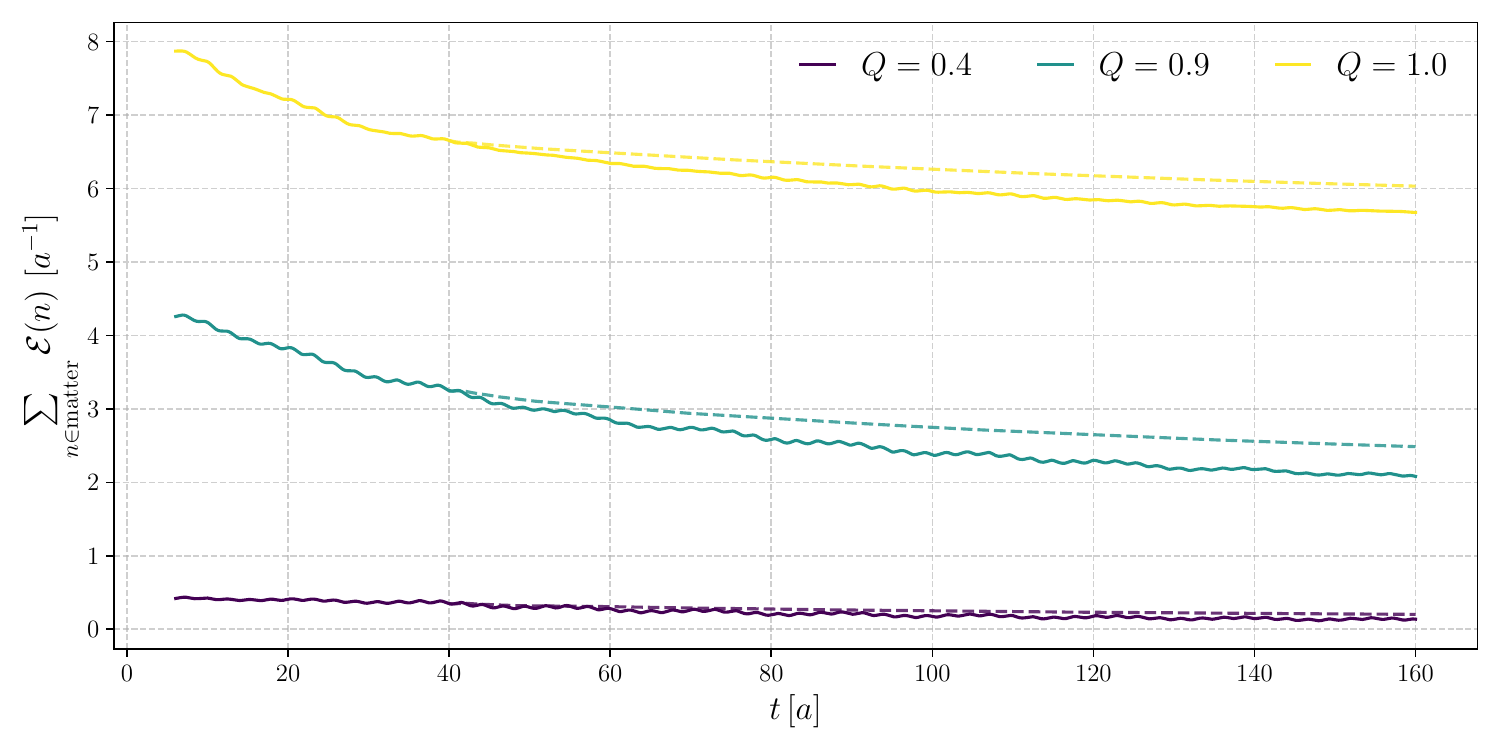}
    \includegraphics[width=.3\textwidth]{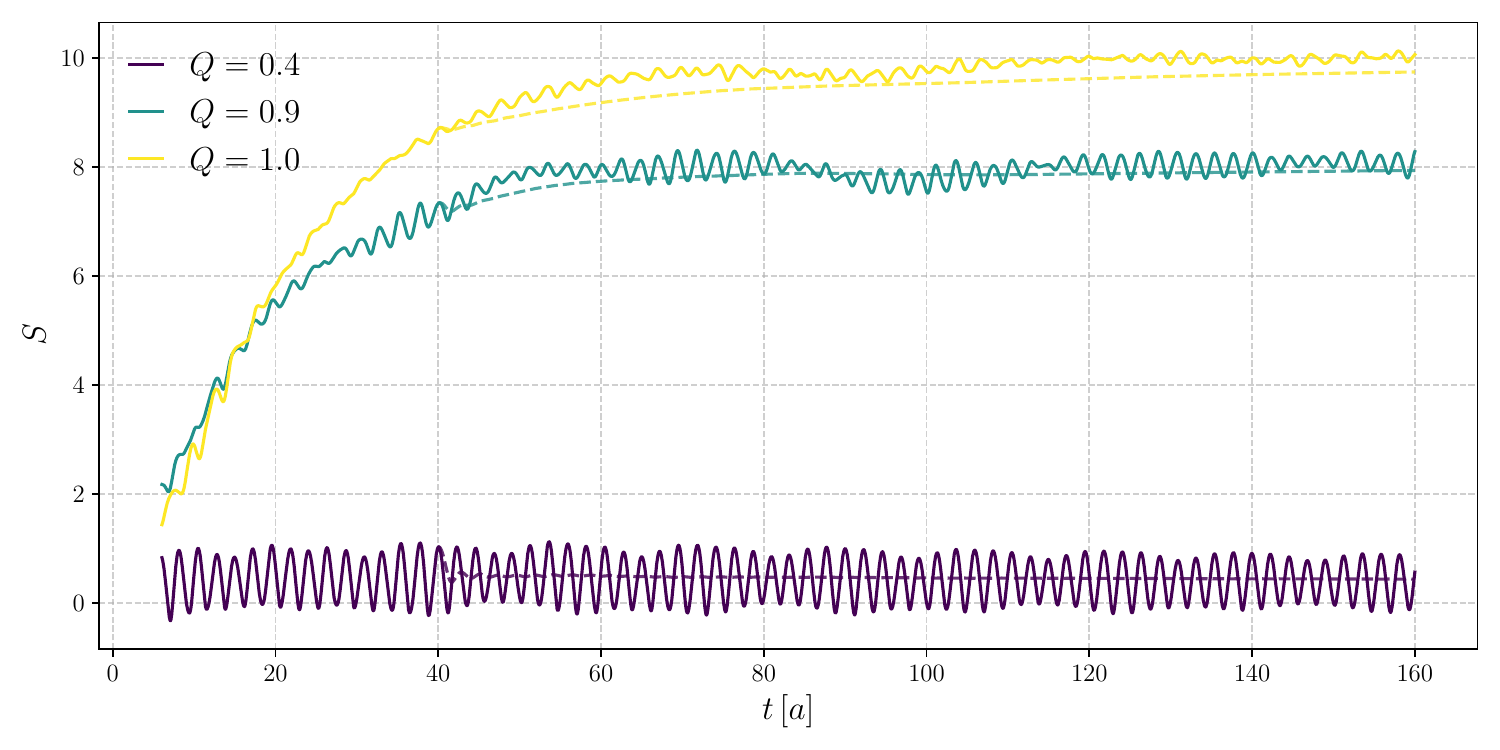}
    \caption{Time evolution of the condensate, energy, and entropy inside the region where the initial electric field is injected. Dashed lines show the late-time average. The matter is assumed to have support between $n=27$ and $n=33$ while increasing this interval does not lead to significant changes.}
    \label{fig:matter_properties}
\end{figure*}


\section{Numerical results}\label{sec:results}

We begin by analyzing the properties of the prepared matter state, computing local observables along its evolution in the absence of the jet state. This is shown in Fig.~\ref{fig:matter_properties}. All the observables asymptotically tend to \textit{equilibrate}, approaching the time-averaged values (dashed lines). This suggests that the prepared matter state should not be too far out of equilibrium, closely resembling e.g. the expected behavior of the QGP phase in heavy ion collisions. Even near the scattering time, $t\approx 60 \, a$, all the expectation values have small variations with time. The evolution of the three observables also has the expected behavior: the local condensate decreases over time as particle density goes down inside the medium, with the energy following the same trend; on the other hand, the entropy increases as the state equilibrates, while at late times it is maximized.

\begin{figure*}
    \centering
    \includegraphics[width=.68\linewidth]{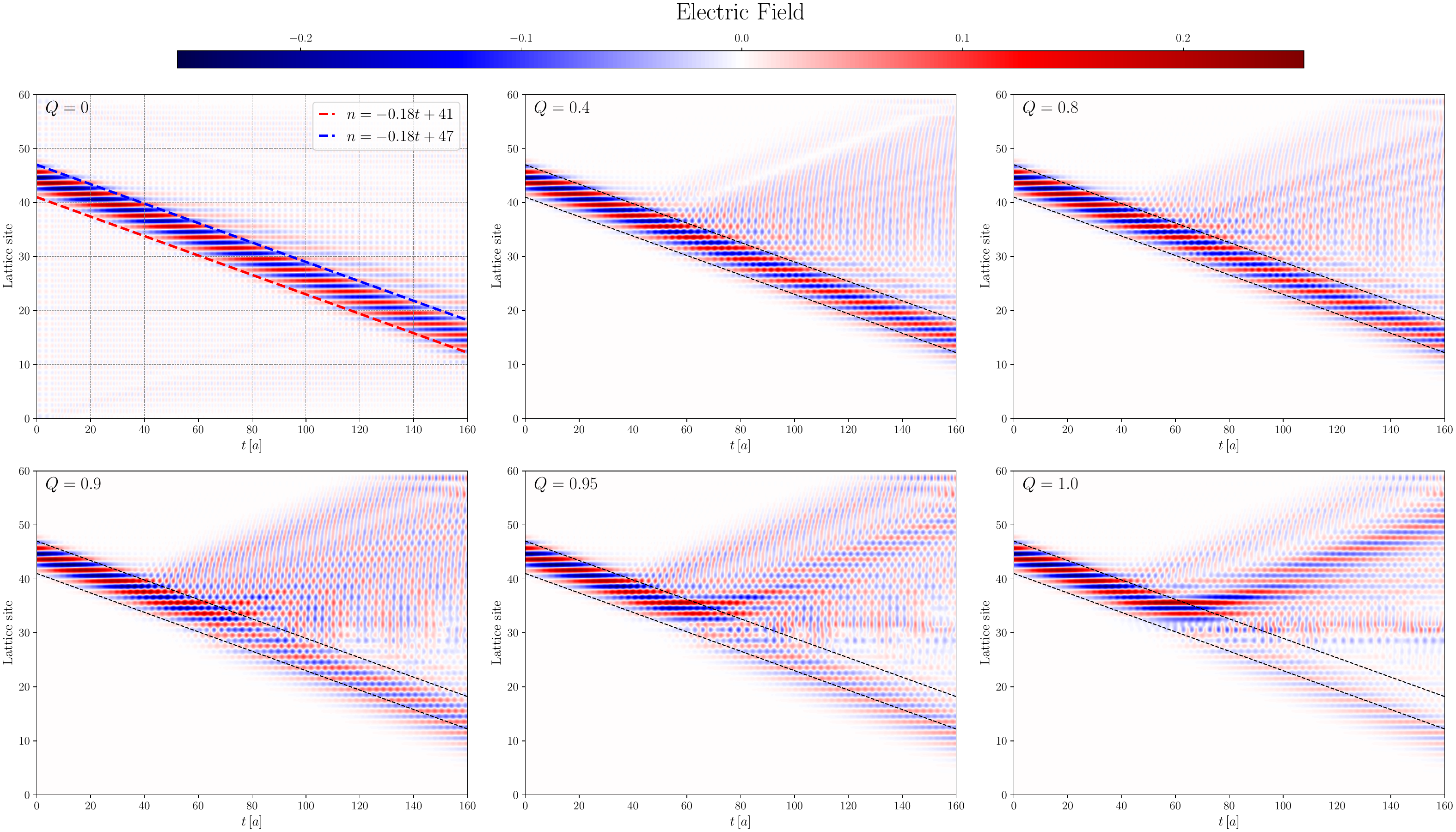}
    \caption{Evolution of the electric field after subtracting the case of a jet state being injected. As a result, before the scattering, all the plots have the same behavior. We plot the classic paths followed by the edges of the initial jet state on the top left figure. The same trajectories are provided on the remaining plots, shown in black.}
    \label{fig:E_field}
\end{figure*}

Moving to the scattering process, in Fig.~\ref{fig:E_field}, we show the time evolution of the electric field on the lattice as a function of time for different values of the external charge $Q$, after removing the contribution from the mater state. Equivalent plots, without subtracting the contributions from the medium, are given in the appendix, along with the dispersion of the electric field values on the lattice.

\begin{figure}[h!]
    \centering
    \includegraphics[width=.8\columnwidth]{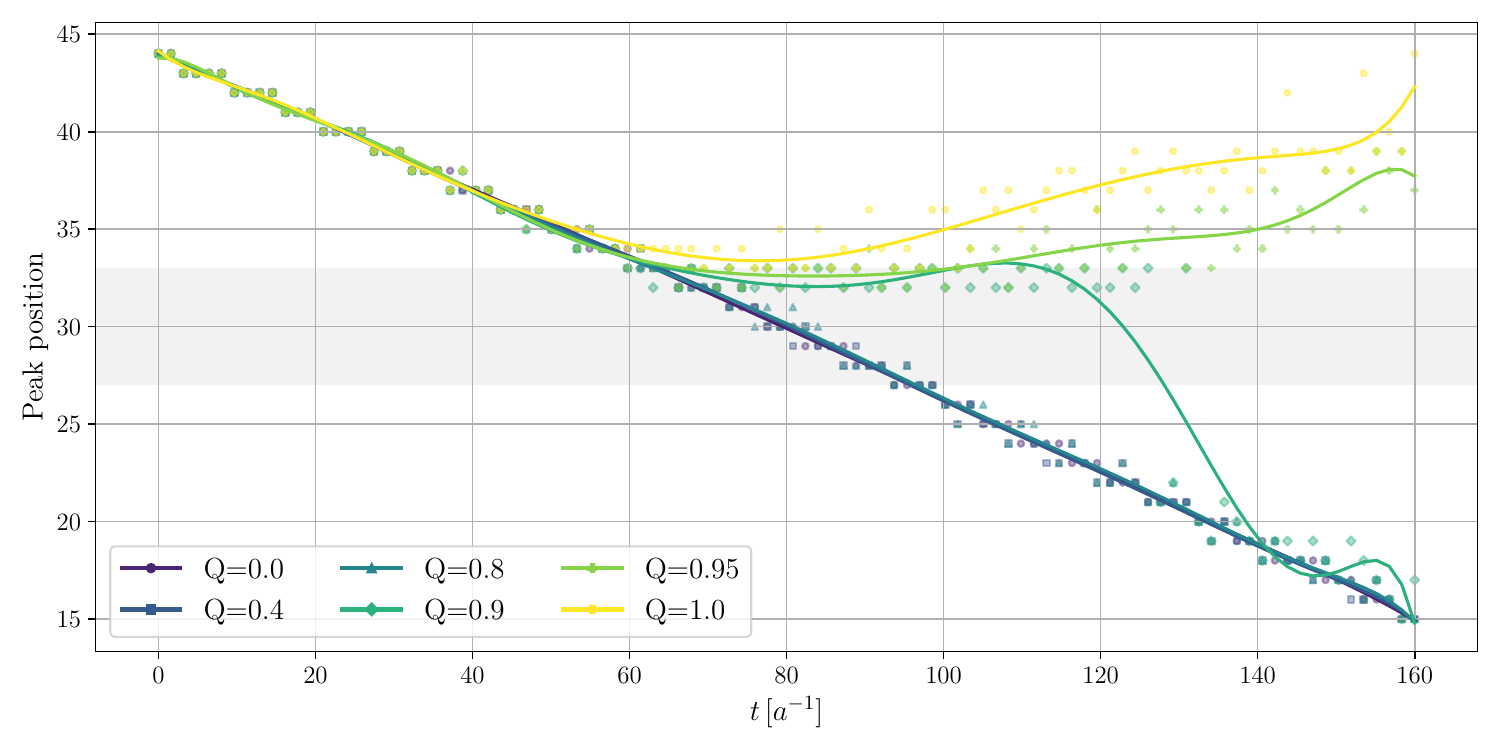}
    \caption{Position of the wave-packet peak for different charges, extracted from the entropy distribution; a qualitatively similar plot can be obtained by extracting the maximum of other observables. Solid curves are obtained by fitting the raw data, and the gray band denotes the initial position of the medium. For the $Q=0.9$ case, the fit changes drastically between the center of the lattice and the right edge, since this is an intermediate step between the probe propagating through the medium and the back-scattering regime.}
    \label{fig:velocity}
\end{figure}

 At small values of $Q$ the jet state propagates nearly ballistically. In the absence of matter, i.e. $Q=0$, the jet evolves with well-defined momentum, and the state does not break apart for the times being considered. We further illustrate this in Fig.~\ref{fig:velocity}, where we plot the peak of the wave-packet as a function of time for several $Q$;\footnote{The maximum is obtained for the entropy distribution, but we checked that a nearly identical picture could be obtained from e.g. the local condensate.} for the lowest ones, the (phase) velocity is almost constant. Nonetheless, as time progresses a small tail of excitations trails the jet, as it locally excites the vacuum. For $0<Q< 0.9$, the picture is slightly modified due to the presence of the medium. Releasing the external charges, the medium is mainly concentrated around the region where it is initially produced. However, part of the medium's energy immediately goes on the light cone, interacting with the meson before it hits the core of the matter. However, the excitations propagating along and inside the light cone do not affect the jet state significantly. Note the stronger the initial $Q$, the more confined the medium is. The interactions with the jet result in the target being locally excited, and some energy being trapped inside the region $27<n <33$, where the initial electric field is inserted. 

For values of $Q\geq0.9$, the picture qualitatively changes. First, we can observe that for $Q=0.9$, the medium excitations due to the passage of the probe are quite significant, and the outgoing jet state is considerably attenuated. Moreover, extracting the semi-classical trajectories from the $Q=0$ case (red and blue dashed lines in Fig.~\ref{fig:E_field} top left plot), we observe that here the outgoing state's trajectory is slightly shifted, resulting from the interaction with the medium. As $Q$ increases this feature becomes more prevalent. Examining Fig.~\ref{fig:velocity}, we see that the distribution peak remains \textit{trapped} in the medium for a significant time for $Q=0.9$, before emerging on the other side of the lattice; for $Q>0.9$ the peak instead moves upwards. We note that the phase shift for $Q\leq 0.9$ is hard to observe in Fig.~\ref{fig:velocity}, due to the fluctuations in the data and the lattice discretization.

For values of $Q>0.9$, we transition from a picture where \textit{most} of the initial energy flux gets transmitted over the medium. Instead, it back-scatters on the target state, see also Fig.~\ref{fig:velocity}. This corresponds to the limit where the probe scatters on a large potential wall, and it is the $1+1$-d analog to the QCD case shown in Fig.~\ref{fig:cartoon} near the so-called black disk limit, where the target structure is unresolved by the probe~\cite{Gribov:1969zz,Bjorken:1973gc}. In this case, it should also be noted that some transmission over the barrier is still visible and that the reflected state is much broader than the incoming one, as observed from its electric field's shape.

Thus, in summary, while increasing the value of the external charges $Q$, we interpolate between a picture where the jet propagates ballistically, to one where it shoots through the medium locally exciting it, to a hard wall limit, where most of the energy flux is reflected.

Having discussed the general picture of the scattering process we next quantify the modifications induced on the jet by the presence of the medium. To this end, we compute the jet energy lost in the medium. In the QCD context, this observable plays a key role in indirectly describing QCD matter from the measured particle spectra, see e.g.~\cite{Apolinario:2022vzg}. In the context of quantum simulation such aspects have been explored in light-front QCD~\cite{Barata:2022wim,Barata:2023clv,Barata:2021yri,Qian:2024gph,Wu:2024adk,Li:2021zaw}, and more recently by examining the behavior of external currents propagating in dense states in the Schwinger model~\cite{Farrell:2024mgu}.

\begin{figure*}
    \centering
    \includegraphics[width=.8\textwidth]{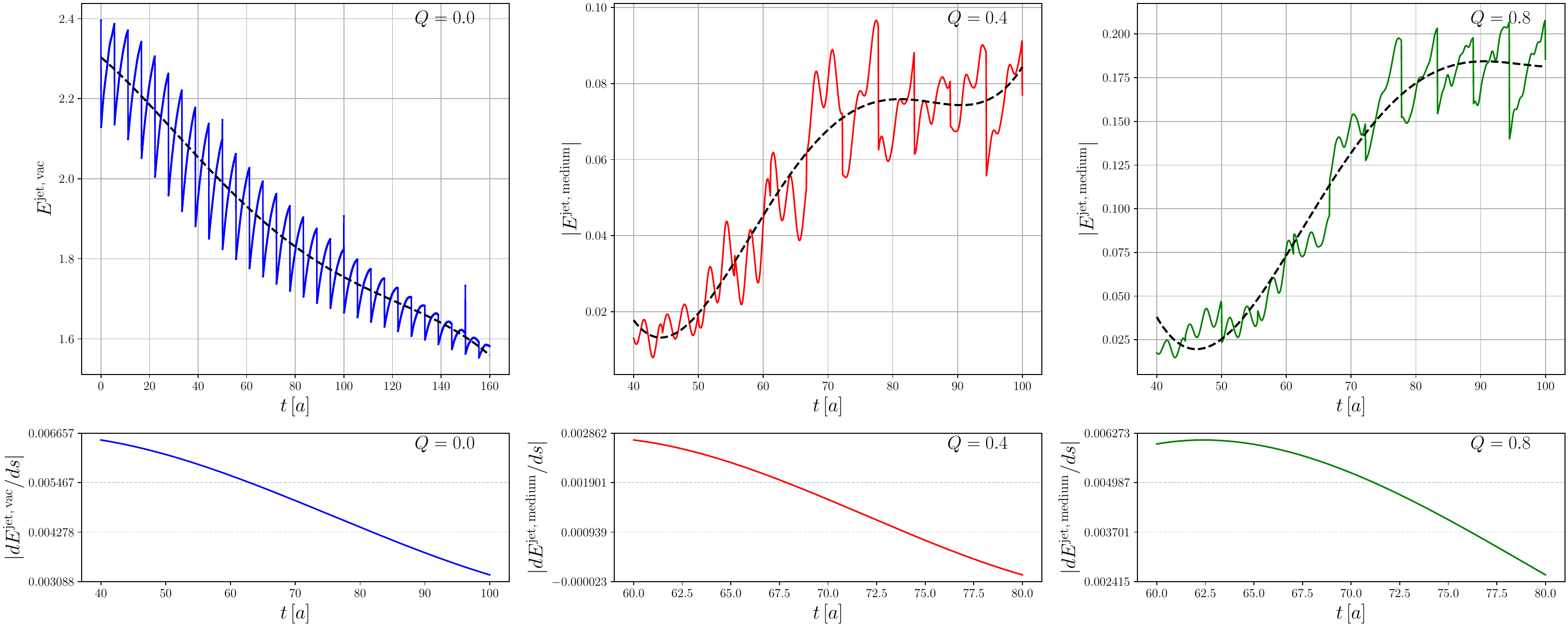}
    \caption{\textbf{Top:} Energy stored inside the jet state for no medium, $Q=0$, and increasing the medium intensity. In dashed black, we show the polynomial fit to the raw lattice data. \textbf{Bottom:} The respective energy loss rate, obtained from the fitting curve.}
    \label{fig:energy}
\end{figure*}

In Fig.~\ref{fig:energy} we show the evolution of the energy stored in the jet state as a function of time for several $Q$, and following the definitions provided in section~\ref{section:model}, where the jet energy is defined using the lattice points in-between the lines drawn in Fig.~\ref{fig:E_field}. Recall from above that this is only reasonable for $Q<0.9$, since after that point the jet mixes with the medium, and most of the energy back-scatters, as shown in Fig.~\ref{fig:velocity}. We note that the definition used is analogous to what occurs in QCD, as illustrated in Fig.~\ref{fig:cartoon} (\textbf{bottom}). In particular, in the vacuum, the energy lost by the jet in QCD is associated with radiation emitted out of the jet cone (Fig.~\ref{fig:cartoon} (\textbf{bottom left})); in the Schwinger model, this $\Delta E$ energy loss is related to the vacuum excitation due to the jet passage, as part of the energy lags. In the case of interactions with the medium, in QCD the jet-medium interactions lead to part of the probe's energy being injected into the matter, which can occur by either elastic or inelastic processes. In the Schwinger model, the picture is similar to the vacuum case, i.e. the probe locally excites the matter state.  

As expected, as the jet propagates its energy gets depleted, even in the vacuum case, as can be directly seen in Fig.~\ref{fig:energy} (\textbf{left}). We note that the oscillatory behavior is a manifestation of the discretization and one should instead consider only the central values. To that end, we performed a polynomial fit to the lattice data. In the cases when $Q\neq 0$, after subtracting the vacuum, there is still a non-zero (and non-fluctuating) evolution, corresponding to the energy depleted into the medium.  Note that the magnitude of the energy loss increases with $Q$, as expected. In the lower panel of Fig.~\ref{fig:energy} we show the energy loss rate, in the relevant time intervals. We observe that the rate has a near-linear dependence on the path length, resembling the findings in QCD for \textit{bremsstrahlung} radiation at high-energies, see e.g.~\cite{Baier:1998kq,Zakharov:1997uu}.

\begin{figure*}[t!]
    \centering
    \includegraphics[width=.7\textwidth]{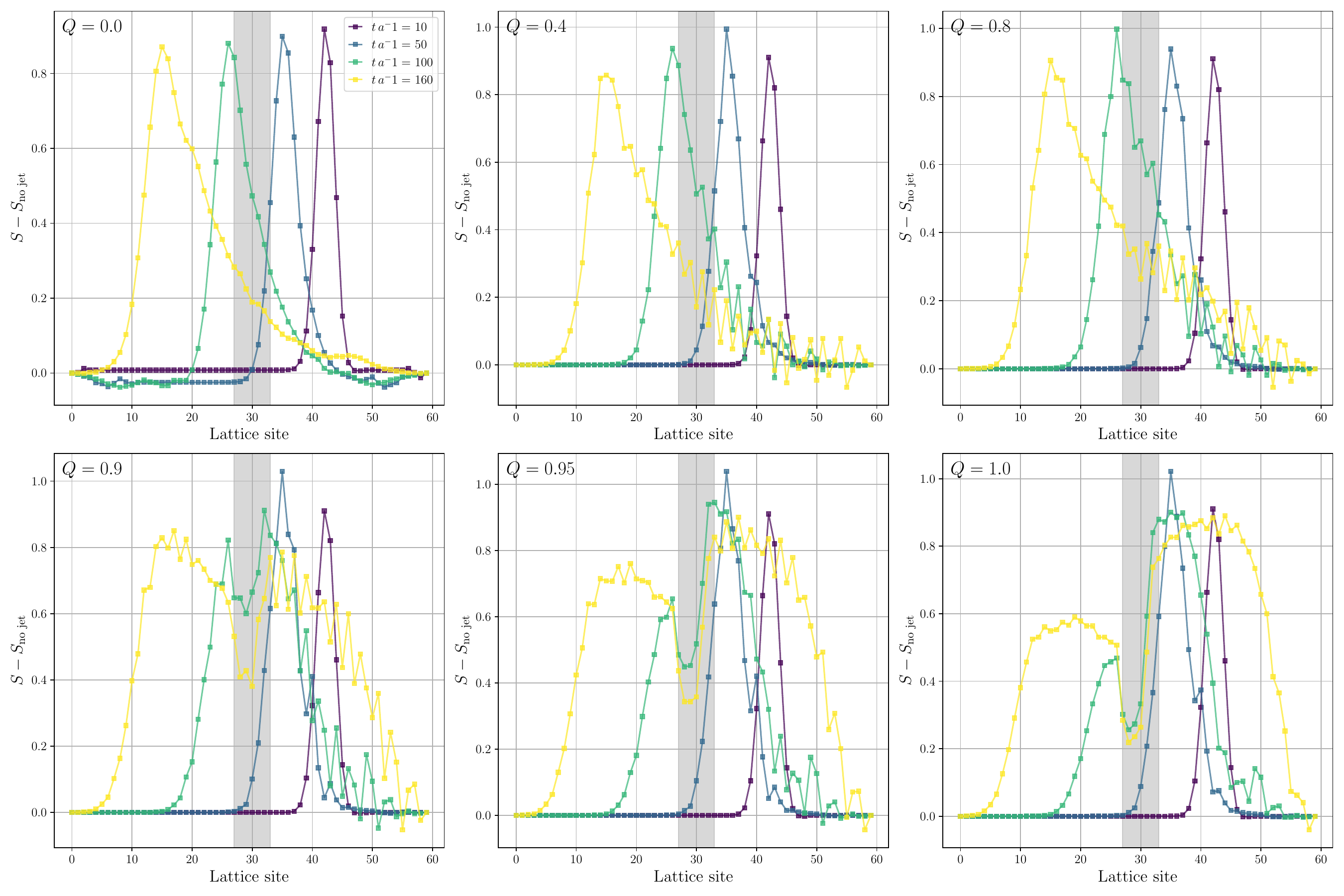}
    \caption{Entropy associated with the propagation of the jet. Here we have subtracted the entropy associated with the medium. Thus all the curves reflect the contributions associated with the jet or the impact of the medium on the probe. The gray band denotes the initial medium location. }
    \label{fig:entropy}
\end{figure*}

Finally, we study the evolution of the entanglement entropy during the collision in Fig.~\ref{fig:entropy}. Previously, this and similar observables had been used as a way to detect inelastic particle production channels in scattering processes involving equal initial states, which is associated with an increase in the entropy, see e.g.~\cite{Pichler:2015yqa,Papaefstathiou:2024zsu,Rigobello:2021fxw}. In contrast, when only elastic scattering processes are at play, the entropy variation should be null. This is observed in Fig.~\ref{fig:entropy} for $Q=0$, where the front of the jet propagates linearly, while a small back tail develops. When $0<Q<0.9$, being an entropy increase in the tail, we still observe a ballistic transport. This entropy excess is associated with the interaction with the medium. At larger values of $Q$, we observe that the entropy peak on the left-hand side significantly decreases, as the state back-scatters on the matter. This is more evident for $Q=1$, where the maximum value of the entropy on the right-hand side of the gray band does not vary compared to the initial condition. However, the final distribution is much broader, indicating that this process is not elastic. This results from the fact that even though the energy flux is deflected backward, its content is altered due to the jet and medium mixing and forming a state where one can no longer distinguish them. 

\begin{figure*}
    \centering
    \includegraphics[width=.68\textwidth]{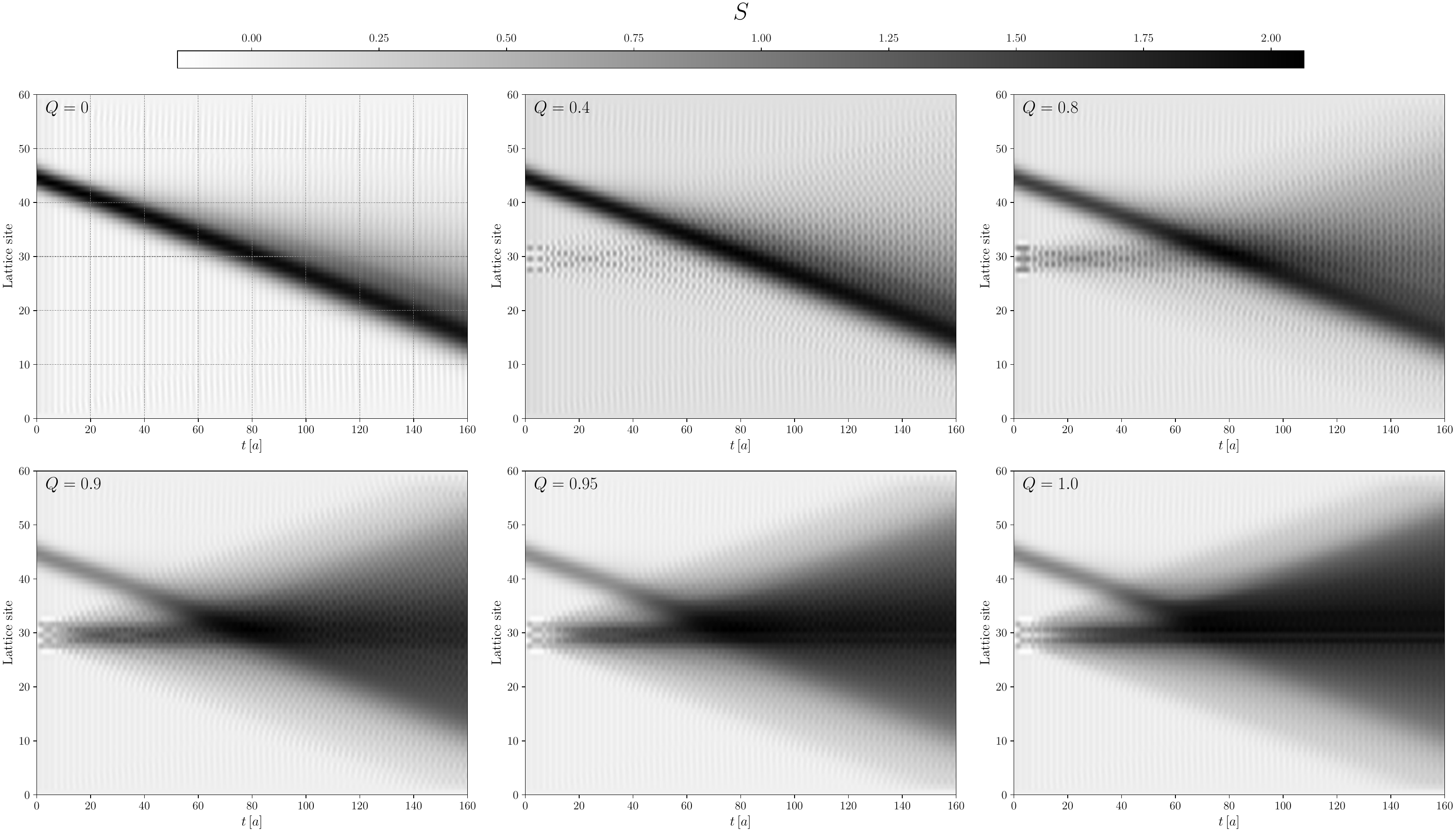}
    \caption{Entropy evolution during the full scattering process for several values of $Q$.}
    \label{fig:entropy_2}
\end{figure*}

To complement these figures, the spacetime evolution of the full entropy is also given in Fig.~\ref{fig:entropy_2}, analogously to the electric field in Fig.~\ref{fig:E_field_2}. Curiously, compared to Fig.~\ref{fig:E_field}, here, for large $Q$, it is much harder to observe the transmitted jet state, and instead one sees that the jet and the medium become indistinguishable. This is expected since the current analysis is observable dependent, and the $Q=0.9$ case is on the edge between the ballistic scenario and the case where the two states mix. These plots make evident that for large $Q$ the jet and the medium form a combined state.
 
\section{Implementation in synthetic platforms}\label{sec:synth_platt}
Here, we briefly discuss the possible implementations of the current protocol in quantum devices. 

The gauge field free form of the Schwinger model, given in Eq.~\eqref{eq:Spin_Hamiltonian}, implies an all-to-all interaction term, which is non-trivial to realize experimentally. Moreover, having higher-dimensional theories in mind, where one can not integrate out the gauge degrees of freedom, it is more reasonable to consider the form of the theory with explicit dependence on the electric field. Since the local Hilbert space of $E$ is still infinite dimensional, we shall focus instead on the related quantum link model (QLM) governed by the Hamiltonian
\begin{align}\label{eq:QLM_Hamiltonian}
	H^{\rm QLM}&= 	\frac{g^2a}{2} \sum_{n=1}^{N-1} L_{z,n}^2  +\sum_{n=1}^{N} m  (-1)^n \chi_n^\dagger  \chi_n \nn
	&-\frac{i}{2a} \sum_{n=1}^{N-1}  \chi^\dagger_n U_+(n) \chi_{n+1} - {\rm h.c.}  \, ,
\end{align}
where the electric field operator $L_{z,n}$ and the link raising operator $U_+(n)$ satisfy the angular momentum algebra, see e.g.~\cite{Chandrasekharan:1996ih,Kogut:1974ag,Zohar:2015hwa,Brower:1997ha}.\footnote{Equivalently we could have chosen to instead use a discreet approximation to the U(1) gauge group, see~\cite{Ercolessi:2017jbi}. Such details are not critical for this short discussion.} The electric field can take up to $2l+1$ values, and when acting on the local link state $|l,m\rangle$ with the raising operator one has 
\begin{align}
   U_+(n)  |l,m\rangle_n = \sqrt{1-\frac{m^2+m}{l^2+l}}|l,m+1\rangle_n
   \, ,
\end{align}
assuming one works in a spin representation sufficiently large to accommodate the highest value of the electric field. This model, which in the limit $l\to \infty$ recovers the Schwinger model, is simple enough to study numerically while avoiding a long-range potential, and it has been implemented in a variety of platforms, see e.g.~\cite{Banuls:2019bmf} for a recent review. 

For the current simulation protocol, while working close to the strong coupling limit, this setup is rather interesting since, as can be seen from the numerical results, the electric field values $|L(n)|<1$. Thus in this case, even a $l=1$ QLM would already offer a reasonable approximation to the U(1) model, see also~\cite{Banuls:2015sta,Papaefstathiou:2024zsu} for related considerations. This case is fairly interesting due to the potential to realize it in a mixed architecture, where the matter states are represented in terms of qubits, while the gauge degrees of freedom are implemented in terms of qutrits (for $l=1$). The experimental realization of such a setup has been studied in depth for the case of $2+1$-d gauge theories~\cite{Meth:2023wzd,Brennen:2015pgn}, see also~\cite{Ale:2024uxf,Araz:2024kkg}.

Our protocol in the QLM formulation has some further advantages. Firstly, the initial state preparation is simpler: one can first prepare the ground state of the theory at non-zero mass, and then inject the matter state by locally exciting the link to a higher spin state while satisfying Gauss's law. Moreover, in this case, one can efficiently modulate the matter in space, allowing for more complex initial states. This procedure does not require any time evolution unlike the MPS simulation above. The preparation of the jet can also be made more realistic while keeping track of the gauge links; for example, one could make a probe where the core is dense (higher electric field values) while the outer structure is dilute. This was partially achieved via the Gaussian profile in Eq.~\eqref{eq:jet_init}. Of course, one could implement similar ideas from the above MPS simulation, but its realization is less straightforward in the fermionic model. 

\section{Conclusion}\label{sec:conclusion}

In this work, we show the first study of the scattering dynamics between a propagating probe and a matter state in the Schwinger model. Compared to high-energy nuclear experiments, this simulation protocol gives a testbed to recreate different setups, such as deep inelastic scattering or high-energy hadronic collisions, in a real-time simulation. Of course, working in a lower dimensional gauge theory, several aspects are qualitatively different from the QCD case. Nonetheless, as a first step in this direction, we believe the current approach can provide novel ways to study matter states, surpassing several experimental limitations. We also note that on the theory side, the current setup goes beyond more traditional frameworks to describe these events, where the medium is treated within some effective description of QCD, and part of the dynamics is lost. In the current case, the matter state is fully dynamic and can lead to, for example, particle production.

From the simulation protocol employed, we could qualitatively identify three distinct scenarios: one where the matter is weak/dilute and the jets propagates ballistically through it; an intermediate regime where the probe still penetrates the medium but part of its energy is depleted; and a more extreme case where the medium acts as a strong wall potential which leads to most of the initial energy flux to back-scatter. Compared to the QCD analog, the first scenario is close to the vacuum evolution of partons/jets; the second case is relevant for understanding phenomena such as jet quenching or heavy flavor transport; the last regime approaches the unitarity bound set by the black-disk limit. Finally, we have further characterized these events by computing the energy depleted from the jet, finding its dependence to be mostly linear with the path length, at the level of the energy rate. Computing the entropy variation during the scattering process, we have found that for large values of $Q$, the jet and the matter combine into a common quantum state.

Looking ahead, it would be interesting to explore similar protocols in $2+1$-d theories, where gauge fields become dynamical. Even in these cases, small-scale simulations of finite dimensional gauge groups would already provide much richer dynamics than the Schwinger model. We have identified that the current protocol could be implemented for a spin-1 QLM while capturing part of the qualitative features seen in the Schwinger model. Detailed implementations of this QLM have been considered in mixed qubit-qutrit architectures in higher dimensions. Another important aspect one would like to improve is the medium's preparation, distinguishing between cold nuclear targets and hot matter. We shall address several of these aspects in forthcoming work.

\textit{Acknowledgments:} We are grateful to Giuliano Giacalone, Adrien Florio, David Frenklakh, Meijian Li, Swagato Mukherjee, Wenyang Qian, Carlos Salgado, Andrey Sadofyev, Enrico Speranza, and Raju Venugopalan for helpful discussions.

\bibliography{refs.bib}

\appendix

\section{Complementary Figures}
\begin{figure}[t!]
    \centering
    \includegraphics[width=.8\columnwidth]{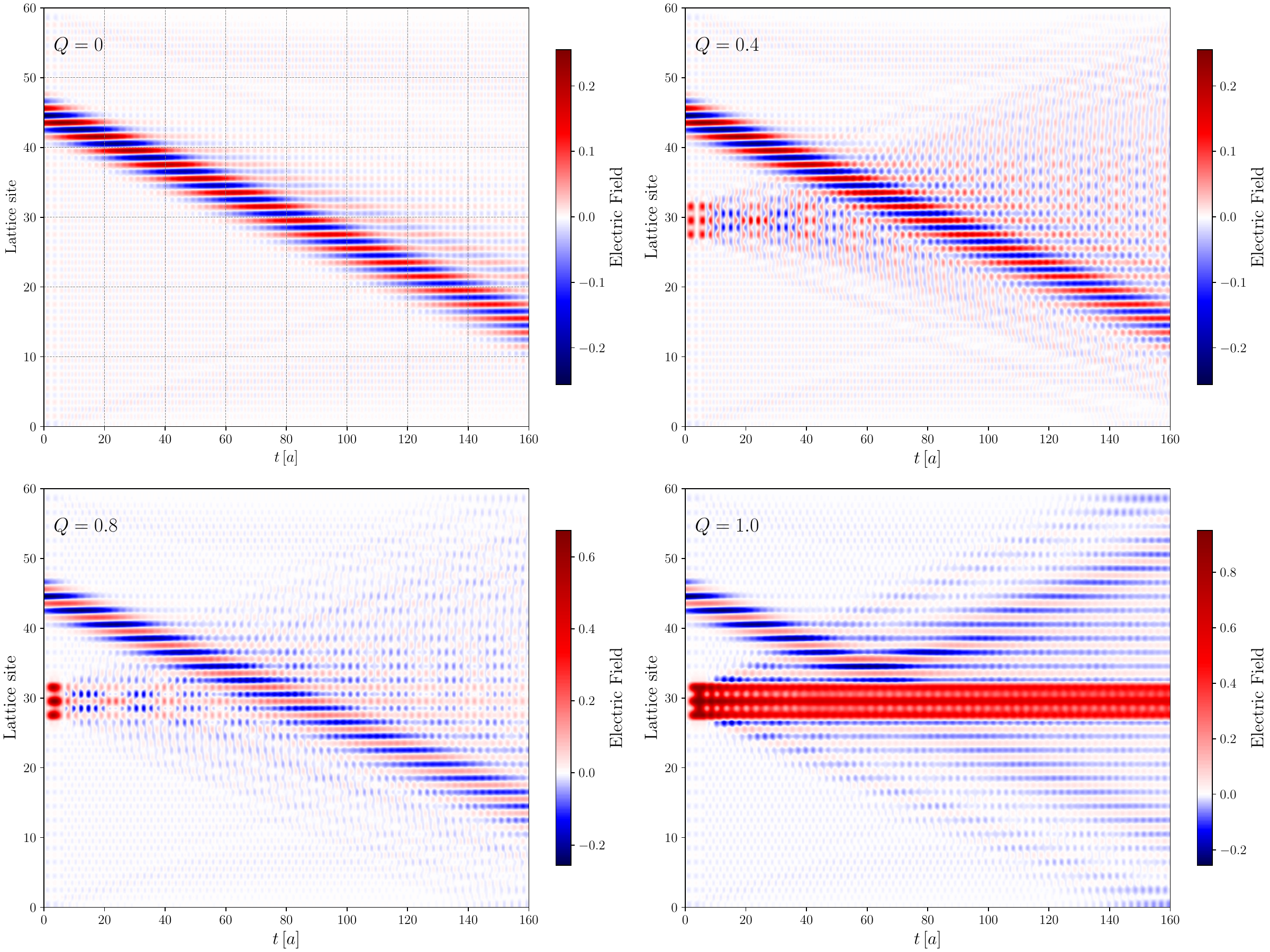}
    \includegraphics[width=.8\columnwidth]{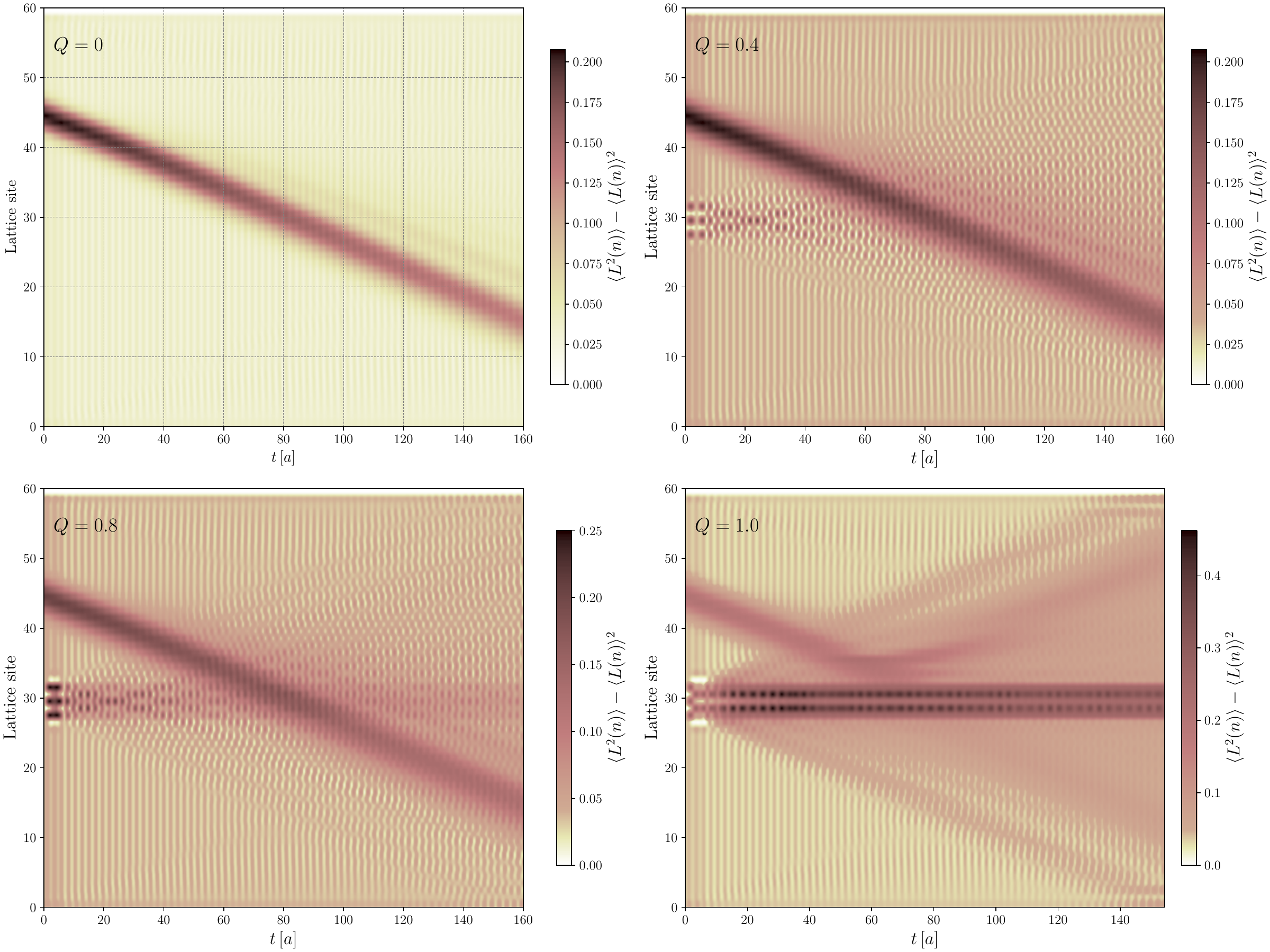}
    \caption{\textbf{Top:} Evolution of the full electric field after subtracting the contributions from the ground state. \textbf{Bottom:} Electric field's dispersion for the same values of $Q$.}
    \label{fig:E_field_2}
\end{figure}

In this appendix, we provide some extra figures, detailing the system evolution studied in the main text. In Fig.~\ref{fig:E_field_2} we show, on top, the full electric field during the system's evolution for several values of $Q$. At the largest values of the initial external charge, we find that the electric field in the matter state is the most intense, nearly reaching its maximum possible value in the strong coupling regime. In the bottom panel, we show the dispersion of the electric field values on the entire lattice, demonstrating their smallness, as expected close to the strong coupling limit.  

\end{document}